\documentclass[aps,prl,floats,floatfix,twocolumn,superscriptaddress,longbibliography]{revtex4-1}
\usepackage{latexsym,amsmath}
\usepackage{amsbsy}
\usepackage[pdftex]{graphicx}
\usepackage{amssymb}
\usepackage{epstopdf}
\usepackage[usenames]{color}
\usepackage{float}
\usepackage{hyperref}
\usepackage{graphicx}
\usepackage{amsmath}
\usepackage[normalem]{ulem}

\newcommand{\sys}{\mathcal{S}}

\begin{document}

\title{Decoherence without entanglement and Quantum Darwinism
}

\author{Guillermo Garc\'ia-P\'erez}
\affiliation{QTF Centre of Excellence, Turku Centre for Quantum Physics, Department of Physics and Astronomy, University
of Turku, FI-20014 Turun Yliopisto, Finland}
\affiliation{Complex Systems Research Group, Department of Mathematics and Statistics,
University of Turku, FI-20014 Turun Yliopisto, Finland}

\author{Dario A.~Chisholm}
\affiliation{Universit\`a degli Studi di Palermo, Dipartimento di Fisica e Chimica -- Emilio Segr\`e,\\ via Archirafi 36, I-90123 Palermo, Italy}

\author{Matteo A.~C.~Rossi}
\affiliation{QTF Centre of Excellence, Turku Centre for Quantum Physics, Department of Physics and Astronomy, University
of Turku, FI-20014 Turun Yliopisto, Finland}

\author{G.~Massimo Palma}
\affiliation{Universit\`a degli Studi di Palermo, Dipartimento di Fisica e Chimica -- Emilio Segr\`e,\\ via Archirafi 36, I-90123 Palermo, Italy}
\affiliation{NEST, Istituto Nanoscienze-CNR, Piazza S.~Silvestro 12, 56127 Pisa, Italy}

\author{Sabrina Maniscalco}
\affiliation{QTF Centre of Excellence, Turku Centre for Quantum Physics, Department of Physics and Astronomy, University
of Turku, FI-20014 Turun Yliopisto, Finland}
\affiliation{QTF Centre of Excellence, Center for Quantum Engineering, Department of Applied Physics,
Aalto University School of Science, FIN-00076 Aalto, Finland}

\begin{abstract}
It is commonly believed that decoherence arises as a result of the entangling interaction between a quantum system and its environment, as a consequence of which the environment effectively measures the system, thus washing away its quantum properties. Moreover, this interaction results in the emergence of a classical objective reality, as described by Quantum Darwinism. In this Letter, we show that the widely believed idea that entanglement is needed for decoherence is imprecise. We propose a new mechanism, dynamical mixing, capable of inducing decoherence dynamics on a system without creating any entanglement with its quantum environment. We illustrate this mechanism with a simple and exactly solvable collision model. Interestingly, we find that Quantum Darwinism does not occur if the system undergoes entanglement-free decoherence and, only when the effect of a super-environment introducing system-environment entanglement is taken into account, the emergence of an objective reality takes place. Our results lead to the unexpected conclusion that system-environment entanglement is not necessary for decoherence or information back-flow, but plays a crucial role in the emergence of an objective reality.
\end{abstract}
\maketitle
{\it Introduction -} The emergence of a classical objective reality from the underlying quantum description of the world is arguably the most studied, debated and still elusive open problem in the foundations of quantum theory. This is known as the quantum measurement problem and it is generally formulated and addressed using the theory of open quantum systems~\cite{Breuer2002,Weiss,RivasHuelgaBook}. The starting point is the realisation that every realistic quantum system is never completely isolated and, therefore, its quantum description must be seen in a more general framework. Specifically, the system of interest is embedded in a larger quantum system, known as its environment. Due to the inevitable interaction with the latter, quantum superpositions are transformed into a classical statistical mixture of the pointer states, which are unaffected by the interaction with the environment \cite{Zurek1981}.  This dynamical phenomenon goes under the name of environment-induced decoherence \cite{Zurek1991,Zurek2003}.

The microscopic description of the system-environment interaction generally allows us to identify the pointer states, but in order to explain how different observers obtain a consistent, and therefore objective, description of reality one must invoke the process known as Quantum Darwinism (QD) \cite{Blume-Kohout2006,Zurek2009}. In words, QD predicts that multiple observers having access to different small fragments of the environment retrieve the same information about the system's state if it is redundantly encoded in such fragments. This is known as objectivity of measurement outcomes and it has been recently experimentally demonstrated in Refs.~\cite{Ciampini2018,Chen2018,Unden2018}. 

The more general concept of objectivity of observables \cite{Horodecki2015} has been demonstrated in Ref.~\cite{Brandao2015} for finite dimensional systems, and in Ref.~\cite{Knott2018} for infinite-dimensional ones, where it was proven that QD is generic, i.e., it occurs independently from the specific model considered (see also Ref.~\cite{Foti2018}). Note that, while the description of decoherence focuses on the dynamics of the open system only---with the environment generally being traced out---QD promotes the role of the environment from passive to active, since it assumes that it is what we actually observe to indirectly retrieve information on the system. Therefore, a dynamical description in terms of the reduced state of the system is not sufficient anymore, and one needs to look at the combined system-environment (or environmental fragments) state instead. 

The very idea that open system dynamics arises from the interaction between two parts of a bipartite total system naturally suggests that entanglement must be established between the two during the time evolution. This is indeed very often the case and it is therefore not surprising that environment-induced decoherence has been until now associated with the presence of entanglement between system $\cal{S}$ and environment $\cal{E}$. However, our results show that, contrarily to this intuition, as long as we limit our attention to the reduced dynamics of the system $\cal{S}$ only, decoherence may take place without entanglement. Specifically, we identify two different microscopic descriptions of the total system, one establishing $\cal{S-E}$ entanglement while the other one not, leading to the same reduced dynamics for $\cal{S}$. 

The same model shows that also information back-flow, recently identified as the source of memory effects, i.e., non-Markovian dynamics \cite{Breuer2009}, does not require $\cal{S-E}$ entanglement. Also here, non-Markovianity is defined by looking at the properties of the dynamical map describing the reduced system. Finally, we show that entanglement plays a pivotal role in the objectification process, since it appears to be needed for it to take place. Our results suggest that, in order to fully grasp the true nature of the quantum-to-classical transition, and in particular to elucidate the role played by entanglement, a description in terms of the open system only may not be sufficient. 

A crucial part of our results is the introduction of a new microscopic collisional model \cite{Scarani2002,Ziman2002,Giovannetti2012,Ciccarello2013,Campbell2018,Lorenzo2017,Lorenzo2016,Filippov2017,Ciccarello2017} allowing us to compute analytically the dynamics not only of the system but also of the system-environment fragments. Within this framework, we propose a new mechanism, dynamical mixing, that can induce decoherence dynamics on a system without creating any entanglement with its environment. The key ingredient is, as its name suggests, a random process that drives the interaction times with the environment. The environment is composed of a set of initially uncorrelated ancillae colliding with the system sequentially and only once, at variance with previous work studying QD in collisional models  \cite{Campbell2019}. This interaction mechanism results on pure dephasing of the system qubit, exhibiting both Markovian and non-Markovian dynamics, depending on the relevant parameters. Our analysis reveals that, while dynamical mixing can give rise to exactly the same qubit dephasing as the one caused by an entangling interaction, it is not capable of accounting for QD. However, the introduction of a super-environment acting as the source for the randomness in the collision times elucidates the origin of QD and the role played by entanglement in it.

\begin{figure}
    \centering
    \includegraphics[width=\columnwidth]{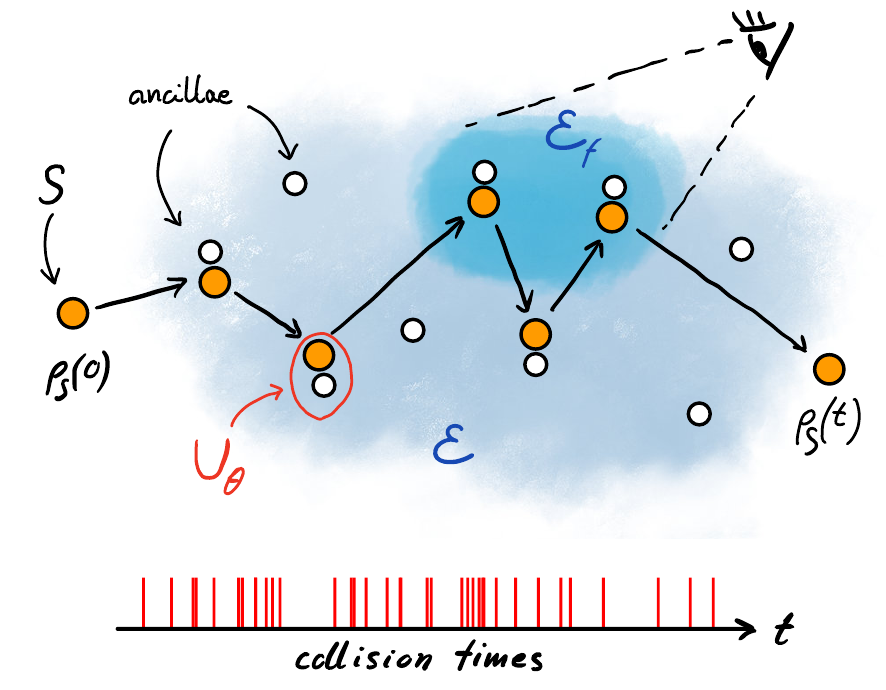}
    \caption{Sketch illustrating the model. The orange dots represent the system $\sys$ colliding with different ancillae (white dots) at exponentially distributed random times, depicted as red vertical lines on the $t$-axis. During each collision, the system and the corresponding ancilla undergo a unitary transformation $U_\theta$. The set of all ancillae defines the environment $\mathcal{E}$, whereas $\mathcal{E}_f$ represents a randomly chosen fraction $f$ of the ancillae, to which an observer might have access in order to acquire information about the state of the system. }
    \label{fig:sketch}
\end{figure}{}
{\it Decoherence without entanglement -} Let us first describe the model under consideration, which is also depicted in Fig.~\ref{fig:sketch}. The system is a single qubit with free Hamiltonian $H_\sys = \frac{\omega}{2} \sigma_z$, where $\omega$ is the qubit frequency, subject to collisions with qubit ancillae at random times. The collision times follow a Poisson process with rate $\lambda$, meaning that the inter-collision time is exponentially distributed. The initial state of all the ancillae is $\rho_a = | 0 \rangle \langle 0 |$, henceforth uncorrelated. When an ancilla collides with the system, it interacts with it with Hamiltonian $H_I = \frac{\eta}{2} \sigma_x^a \otimes \sigma_z^\sys$, where the superscripts stand for ancilla and system, respectively. As it is customary in collisional models, the interaction time is considered to be extremely short, so the collision can be regarded as instantaneous, and its effect amounts to a unitary transformation applied to both the system and the ancilla. Here we denote by interaction strength the limit $\theta = \lim_{t \to 0} t \eta$, where $t$ is the duration of the collision. The resulting unitary transformation for the collision is $U_\theta = e^{-i \frac{\theta}{2} \sigma_x^a \otimes \sigma_z^\sys}$. 

Since all the ancillae are initially uncorrelated with each other and with the system, and they collide with the system only once, we can describe the change on the system's state in terms of the collision channel $\Phi_c [ \rho_\sys ] = \mathrm{Tr}_a \left[ U_\theta \rho_a \otimes \rho_\sys U_\theta^\dagger \right]$.
In the eigenbasis of the interaction Hamiltonian $H_I$, the collision channel can be cast in Kraus form as $\Phi_c [ \rho_\sys ] = K \rho_\sys K^\dagger + K^\dagger \rho_\sys K$, with
\begin{equation}
K = \frac{1}{\sqrt{2}} \begin{pmatrix}
e^{-i \theta/2} & 0\\
0 & e^{i \theta/2}
\end{pmatrix}.
\end{equation}
As a result, the effect of the channel is a factor $\cos \theta$ multiplying the coherences of the qubit state (see Supplemental Material for details). The state of the system at time $t$ is given by the convex sum of all possible stochastic realisations of the ancillary dynamics (trajectories), each of them weighted by its probability.
As shown in the Supplemental Material,  the dynamics of the state of the system is described by the exact master equation
\begin{equation}\label{eq:master_equation}
\dot{\rho}_\sys(t) = -i \left[ H_\sys, \rho_\sys(t) \right] + \lambda \left( \Phi_c \left[ \rho_\sys(t) \right] - \rho_\sys(t) \right),
\end{equation}
which is in Gorini-Kossakowski-Sudarshan-Lindblad form, with $K$ and $K^\dagger$ the Lindblad operators  \cite{Breuer2002}. Hence, we conclude that the system undergoes Markovian dynamics. 

The master equation \eqref{eq:master_equation} describes a pure dephasing dynamics with decoherence factor $c(t) = \exp[-\lambda(1-\cos \theta ) t]$. We notice that $c(t)$ is invariant with respect to changes of the interaction strength $\theta$ upon a proper modification of the collision rate. In particular, the system undergoes the same temporal evolution for
$ \lambda \left( 1 - \cos \theta \right) = C $, where $C$ is constant.
This is an interesting observation, since the interaction strength $\theta$ regulates the level of entanglement between the system and a given ancilla after a collision has taken place. For instance, for $\theta = ( 2m + 1) \pi, \, m \in \mathbb{Z}$, $U_\theta | 0\rangle_a \otimes | + \rangle_\sys = e^{-i \pi/2}| 1\rangle_a \otimes | - \rangle_\sys$, yielding a product state, while for $\theta = ( 2m + 1) \pi/2, \, m \in \mathbb{Z}$, the two become maximally entangled. Hence, we can conclude that, in the former case, the system undergoes pure dephasing while remaining in a separable state with respect to the environment.

The source of randomness in the collision times can be seen as originating from a quantum process where the particle is emitted, for instance, as a result of a spontaneous emission process. This would reintroduce entanglement, in this case with an effective super-environment, in the overall picture. We will analyse the consequences of such an effective description in more detail when focusing on QD. At this point, it is sufficient to stress that a quantum super-environment does not need to enter the description, since the collisions with the ancillae can be triggered by some classical and largely macroscopic stochastic process.

\begin{figure}
\centering
\includegraphics[width=\columnwidth]{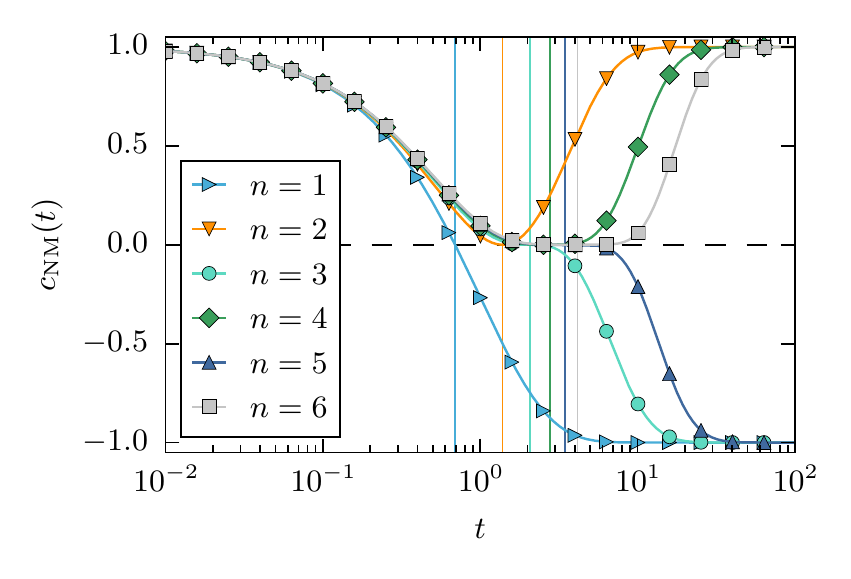}
\caption{\label{fig:non-markovianity} Coherence factor for the qubit, Eq.~\eqref{eq:Non-mark_coh}, as a function of time for different number of ancillae with non-entangling interaction strength, $\theta = \pi$ and $\lambda = 1$. The vertical lines indicate the corresponding mixture time $t_m$, at which the coherence vanishes. As the number of ancillae increases, the coherence remains close to zero for longer periods of time. Notice that the final state depends on the parity of the number of ancillae. }
\end{figure}

{\it Non-Markovianity with fresh ancillae - }The model introduced above describes the situation in which the system undergoes Markovian pure dephasing dynamics while remaining in a separable state with the environment. We now show that dynamical mixing can induce non-Markovian behaviour as well. To this end, we modify the previous model by limiting the number of ancillae to a finite amount $n$. Here, every ancilla's collision time has an exponential probability density with rate $\lambda/n$, while the effect of the collisions is not altered. The integrated dynamics reduces to that of infinitely many ancillae at short times, $\lambda t \ll n$~\footnote{This stems from the fact that the number of collisions at time $t$ follows a binomial distribution with mean $n p_t$, where $p_t = 1 - e^{-\lambda t/n}$ is the probability for a given ancilla to have collided with the system at time $t$.}. In the integrated dynamics, coherences are multiplied by the factor
\begin{equation}
    \label{eq:Non-mark_coh}
    c_{\mathrm{NM}}(t)=\left[ 1 + \left( \cos \theta -1 \right) \left(1 - e^{-\lambda t/n} \right) \right]^n.
\end{equation}
In the particular case in which the system and the ancilla entangle maximally after a collision ($\cos \theta = 0$), the system dephases monotonically with $c_{\mathrm{NM}}(t) = e^{- \lambda t}$, exactly like in the model with infinitely many ancillae. In the case of entanglement-free interaction ($\cos \theta = -1$), however, the off-diagonal elements of the density matrix are multiplied by the factor $c_{\mathrm{NM}}(t) = \left( 2 e^{-\lambda t/n} - 1 \right)^n$, which is equal to zero at $t_m = n \ln 2 / \lambda$ (mixture time) and tends to $(-1)^n$ as $t \to \infty$. As a consequence, if the initial state of the system is, e.g., $\rho_\sys(0) = |+ \rangle \langle + |$, it becomes maximally mixed at $t = t_m$ and it gradually recovers its purity thereafter. Moreover, the system remains in a highly mixed state for longer periods as the environment size $n$ increases (see Fig.~\ref{fig:non-markovianity}). Remarkably, this phenomenon of recoherence takes place despite the fact that the system never collides with the same ancilla more than once, and despite the absence of interactions \cite{Ciccarello2013,McCloskey2014} or initial correlations \cite{Filippov2017} between ancillae.

{\it Quantum Darwinism - }Our results so far reveal that it is possible for a system to undergo exactly the same decoherence dynamics whether or not it becomes entangled with its environment, or even to exhibit non-Markovian dynamics, as a consequence of dynamical mixing. Needless to say, this raises the question of what role does entanglement play in the quantum-to-classical transition. In what follows, we address this issue in the context of QD. As we will show, decoherence without entanglement does not allow for the encoding of information about the system's state into the environment, whereas this is possible when one considers a super-environment giving a quantum origin to the randomness in the collision times. 

QD explains the emergence of objective reality through the mutual information between system and environment. In particular, if several observers that measure different parts of the environment gather the same information about the state of the system, they can consider such information as objective reality. For that to happen, however, there must be some redundancy in how that information is distributed across the environment, and the typical way to quantify it is by calculating the mutual information $I_f$ between the system and a randomly chosen fraction $\mathcal{E}_f$ of the environment, as a function of the fraction's size $f$. Such curve reveals the presence of objective reality through a plateau that spans over a wide range of environmental fraction sizes, and whose value is approximately equal to the von Neumann entropy of the system. In order to assess whether this phenomenon is present in the model introduced in this Letter, we need to calculate $I_f = H_\mathcal{S} + H_{\mathcal{E}_f} - H_{\mathcal{S}\mathcal{E}_f}$, where  $H$ stands for the von Neumann entropy.

\begin{figure}
\centering
\includegraphics[width=\columnwidth]{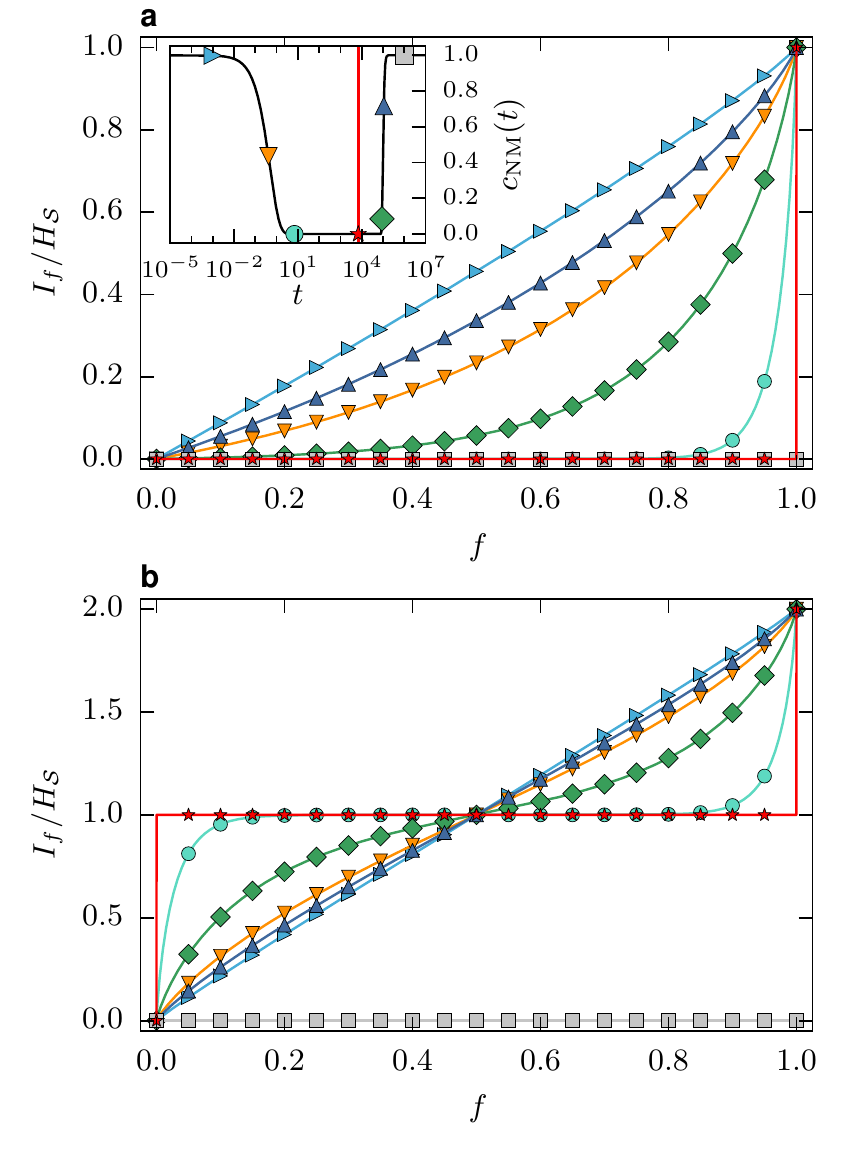}
\caption{\label{fig:Darwinism} Mutual information over system entropy as a function of environment fraction for the two settings at different times, for $n = 10^4$ ancillae and $\lambda = 1$. To indicate the state of the system at the time to which each curve corresponds, they have been coloured matching the corresponding dot in the inset, which shows the coherence factor as in Fig.~\ref{fig:non-markovianity}. The times have been chosen to cover the different dynamical regimes undergone by the system.~\textbf{a}, the origin of the randomness in the collision times is not considered. The plateau in the mutual information occurs for $I_f = 0$, meaning that the ancillae alone barely carry any information.~\textbf{b}, the emitters are part of the quantum state as well. The effect of the emitters is the appearance of QD while the system is in a highly mixed state. }
\end{figure}

We focus first on the case in which the interaction is non-entangling, namely $\theta = \pi$, and the initial state of the system is $\rho_\mathcal{S}(0) = | + \rangle \langle + |$. By tracing out $k = (1-f)n$ ancillae from the total state of system and environment $\rho_{\mathcal{S}\mathcal{E}}(t)$, one can calculate the reduced state when only a fraction $f$ of the latter is considered, $\rho_{\mathcal{S}\mathcal{E}_f}(t) = \mathrm{Tr}_{k} \left[ \rho_{\mathcal{S}\mathcal{E}}(t) \right]$, while further tracing out the system $\mathcal{S}$ yields the reduced state of the fraction of the environment, $\rho_{\mathcal{E}_f}(t)$. The simplicity of our model allows us to perform these calculations analytically, furthermore resulting in a density operators in diagonal form, from which computing their von Neumann entropy is straightforward. All the details of the calculations are given in the Supplemental Material. The resulting mutual information is
\begin{equation}\label{eq:mutual_information_ancillae}
I_f = H_{\mathrm{b}} (P_{n}^{\mathrm{e}}(t)) - H_{\mathrm{b}} (P_{k}^{\mathrm{e}}(t))
\end{equation}
where $f = 1-\frac{k}{n}$, $H_\mathrm{b}(x) = - x \log x - (1-x) \log (1-x) $ is the binary entropy function and $P_{m}^{\mathrm{e}}(t)$ is the probability for $m$ ancillae to yield an even number of collisions at time $t$. This quantity can be computed exactly and reads
\begin{equation}
P_{m}^{\mathrm{e}}(t) = \frac{1}{2} \left[1 + \left( 2e^{-\lambda t/n} -1 \right)^{m} \right].
\end{equation}
In Fig.~\ref{fig:Darwinism}\textbf{a}, we show this curve for different dynamical regimes. Despite an almost linear dependence in some periods, it is mostly flat around null mutual information except for $f \approx 1$ when the system is highly mixed, which implies the absence of objective reality upon which observers can agree.

We now study the case in which the ancillae are emitted as a consequence of a quantum process. In particular, we consider $n$ emitters initially excited that relax to their ground state emitting an ancilla in the process. Moreover, we further assume that the ancilla is emitted in state $\rho_a = | 0 \rangle \langle 0 |$ and, once emitted, it immediately collides with the system, flipping its state (since $\theta = \pi$). Hence, the emitter-ancilla dynamics is such that their joint state at time $t$ can be written as $\sqrt{e^{-\lambda t/n}} |1\rangle_e \otimes |0\rangle_a + \sqrt{1 - e^{-\lambda t/n}} |0\rangle_e \otimes |1\rangle_a$. In this new setting, tracing out the emitters yields exactly the same reduced state as is the previous situation. However, when considering as environment fraction $\mathcal{E}_f$  pairs comprising both the emitter and the corresponding ancilla, the mutual information takes the form
\begin{equation}\label{eq:mutual_information_emitters}
I_f = H_{\mathrm{b}} (P_{n}^{\mathrm{e}}(t)) + H_{\mathrm{b}} (P_{n-k}^{\mathrm{e}}(t)) - H_{\mathrm{b}} (P_{k}^{\mathrm{e}}(t)).
\end{equation}
Comparing this result with Eq.~\eqref{eq:mutual_information_ancillae}, we see that the presence of the emitters introduces the term $H_{\mathrm{b}} (P_{n-k}^{\mathrm{e}}(t))$, which corresponds to the entropy of the environment, $H_{\mathcal{E}_f}$ (see SM). Figure \ref{fig:Darwinism}\textbf{b} depicts the mutual information in this new setting. In this case, there is a clear plateau at $I_f/H_\sys = 1$, revealing a structure of the total state of system and environment compatible with QD. The fact that the plateau is only present at times in which the reduced state of the system is highly mixed, along with our previous discussion regarding the separability of the system and the ancillae alone, can be interpreted as further evidence that the emergence of objective reality requires entanglement.

{\it Conclusions - }
We have investigated the emergence of classical reality from its quantum substrate by means of an exact collisional model, for which we have derived analytically not only the master equation and the dynamical map, but also the relevant system-environment dynamical properties. The stochastic element introduced in the microscopic description of the collisions lends itself to an interesting generalization in terms of a super-environment that keeps track of the occurrence of the collisions. Due to these features, the underlying dynamics is dominated by  simple fundamental physical mechanisms allowing us to shed new light on the role of quantum entanglement in three crucial phenomena: decoherence, non-Markovianity as information back-flow, and Quantum Darwinism. The study of this model leads to the unexpected conclusion that system-environment entanglement is not necessary for decoherence or information back-flow, but plays a crucial role in the emergence of an objective reality.

\begin{acknowledgments}
G.G.-P., M.A.C.R. and S.M.~acknowledge financial support from the Academy of Finland via the Centre of Excellence program (Project no.~312058 as well as Project no.~287750). G.M.P.~acknowledges  PRIN project 2017SRNBRK QUSHIP funded by MIUR. G.G.-P.~acknowledges support from the emmy.network foundation under the aegis of the Fondation de Luxembourg.
\end{acknowledgments}

\bibliography{bibliography}

\onecolumngrid
\clearpage

\renewcommand{\theequation}{\thesubsection S.\arabic{equation}}
\setcounter{equation}{0}
\def\Eqf{{8}}

\section*{\large{Supplemental Material}}

\section{System dynamics}
In this section, we derive the master equation describing the evolution of the system when the number of ancillae is infinite, along with the dynamical map in this case as well as for a finite amount of them.
\subsection{Master equation}
To derive the master equation, we start by assessing the effect of a single collision on the state of the system. Given that all the ancillae are initially uncorrelated---and not correlated with the system either---and they collide only once with the system, we can describe the change on the system's state in terms of the collision channel
\begin{equation}\label{eq:coll_chan}
\Phi_c [ \rho_\sys ] = \mathrm{Tr}_a \left[ U_\theta \rho_a \otimes \rho_\sys U_\theta^\dagger \right].
\end{equation}
In the eigenbasis of the interaction Hamiltonian $H_I$, $U_\theta = \mathrm{diag}\left( e^{- i \theta/2}, e^{i \theta/2}, e^{i \theta/2}, e^{-i \theta/2} \right)$ and
\begin{equation}
\rho_a \otimes \rho_\sys = \frac{1}{2} \begin{pmatrix} \rho_\sys & \rho_\sys \\ \rho_\sys & \rho_\sys \end{pmatrix}.
\end{equation}
Hence, from Eq.~\eqref{eq:coll_chan}, we see that the collision channel in Kraus form reads $\Phi_c [ \rho_\sys ] = K \rho_\sys K^\dagger + K^\dagger \rho_\sys K$, with
\begin{equation}
K = \frac{1}{\sqrt{2}} \begin{pmatrix}
e^{-i \theta/2} & 0\\
0 & e^{i \theta/2}
\end{pmatrix}.
\end{equation} 
As a result, the effect of the channel is a factor $\cos \theta$ multiplying the coherences of the qubit state.

We notice that $K$ commutes with $H_\sys$ and, therefore, with $U_t^\sys = e^{- i t H_\sys}$. This in turn implies that, if the number of collisions at time $t$ is $k$, we can simply write the state of system as $\rho_k (t) = U_t^\sys \Phi_c^{(k)} \left[  \rho_\sys (0) \right] U_t^{\sys\dagger}$, where the superscript $(k)$ indicates that the channel is applied $k$ times. Now, the state of the system at time $t$ is given by the convex sum of all possible stochastic realisations of the ancillary dynamics (trajectories), each of them weighted by its probability. According to our previous observation, the state in every trajectory is fully determined by the number of collisions, so we can write
\begin{equation}
\rho_\sys (t) = \sum \limits_{k=0}^{\infty} p_k (t)  \rho_k(t),
\end{equation}
where $p_k(t)$ is the probability for $k$ collision events to have happened at time $t$. Since collisions follow a Poisson process,
\begin{equation}\label{eq:convex_sum_over_num_colls}
\rho_\sys (t) = e^{-\lambda t} \sum \limits_{k=0}^{\infty} \frac{(\lambda t)^{k}}{k!} \rho_k(t).
\end{equation}
The corresponding time derivative hence gives
\begin{equation}
\dot{\rho}_\sys(t) = - \lambda \rho_\sys (t) + e^{-\lambda t} \left( \sum \limits_{k=0}^{\infty} \frac{(\lambda t)^{k}}{k!} \dot{\rho}_k(t)  + \sum \limits_{k=1}^{\infty} \frac{\lambda^k t^{k-1}}{(k-1)!} \rho_k(t)\right).
\end{equation}
The leftmost term in the parenthesis gives the unitary evolution of the system,
\begin{equation}
\begin{aligned}
e^{-\lambda t} \sum \limits_{k=0}^{\infty} \frac{(\lambda t)^{k}}{k!} \dot{\rho}_k(t) &= e^{-\lambda t} \sum \limits_{k=0}^{\infty} \frac{(\lambda t)^{k}}{k!} \left( -i \left[ H_\sys, \rho_k(t) \right] \right) = -i \left[ H_\sys, \rho_\sys (t) \right],
\end{aligned}
\end{equation}
whereas the rightmost one is
\begin{equation}
\begin{aligned}
\lambda e^{-\lambda t} \sum \limits_{k=0}^{\infty} \frac{\lambda^k t^{k}}{k!} \rho_{k+1}(t) &= \lambda e^{-\lambda t} \sum \limits_{k=0}^{\infty} \frac{\lambda^k t^{k}}{k!} \Phi_c \left[ \rho_{k}(t) \right] = \lambda \Phi_c \left[ \rho_\sys (t) \right].
\end{aligned}
\end{equation}
Finally, we can write
\begin{equation}
\dot{\rho}_\sys(t) = -i \left[ H_\sys, \rho_\sys (t) \right] + \lambda \left( \Phi_c \left[ \rho_\sys (t) \right] - \rho_\sys (t) \right),
\end{equation}
which is in GKSL form, with $K$ and $K^\dagger$ as Lindblad operators. Notice that no approximations have been made in the derivation. Hence, we conclude that the system undergoes Markovian dynamics.

\subsection{Integrated dynamics}
Invoking again the commutativity of the Kraus operators with the system Hamiltonian, and using Eq.~\eqref{eq:convex_sum_over_num_colls}, the state of the system at time $t$ can be expressed as
\begin{equation}\label{eq:rho_s_t_coll}
\rho_\sys (t) = e^{-\lambda t} \sum \limits_{k=0}^{\infty} \frac{(\lambda t)^{k}}{k!} \Phi_c^{(k)} \left[  U_t^\sys \rho_\sys (0) U_t^{\sys\dagger} \right].
\end{equation}
Denoting
\begin{equation}
\rho_\sys (0) = \begin{pmatrix}
\rho_{00} & \rho_{01} \\
\rho_{10} & \rho_{11}
\end{pmatrix},
\end{equation}
the term $U_t^\sys \rho_\sys (0) U_t^{\sys\dagger}$ becomes
\begin{equation}
U_t^\sys \rho_\sys (0) U_t^{\sys\dagger} = \begin{pmatrix}
\rho_{00} & \rho_{01} e^{-i \omega t} \\
\rho_{10} e^{i \omega t} & \rho_{11}
\end{pmatrix}.
\end{equation}
Therefore, using the Kraus decomposition of the collision channel, we see that
\begin{equation}\label{eq:k_collisions}
\Phi_c^{(k)} \left[  U_t^\sys \rho_\sys (0) U_t^{\sys\dagger} \right] = \begin{pmatrix}
\rho_{00} & \rho_{01} e^{-i \omega t} \cos^k \theta\\
\rho_{10} e^{i \omega t} \cos^k \theta & \rho_{11}
\end{pmatrix}.
\end{equation}
Introducing this result into Eq.~\eqref{eq:rho_s_t_coll}, we can write
\begin{equation}
\rho_\sys (t) = \begin{pmatrix}
\rho_{00} & \rho_{01} e^{-i \omega t} \langle \cos^k \theta \rangle_{t}\\
\rho_{10} e^{i \omega t} \langle \cos^k \theta \rangle_{t} & \rho_{11}
\end{pmatrix},
\end{equation}
where $\langle \cos^k \theta \rangle_{t}$ is a shorthand notation for
\begin{equation}
\begin{aligned}
\langle \cos^k \theta \rangle_{t} &= e^{-\lambda t} \sum \limits_{k=0}^{\infty} \frac{(\lambda t)^{k}}{k!} \cos^k \theta = e^{-\lambda t} e^{\lambda t \cos \theta} = e^{-\lambda t \left( 1 - \cos \theta \right)}
\end{aligned}
\end{equation}
and yields the coherence factor $c(t)$ in the main text. Hence, the state of the system at time $t$ is
\begin{equation}\label{eq:dyn_map}
\rho_\sys (t) = \begin{pmatrix}
\rho_{00} & \rho_{01} e^{- t \left(i \omega + \lambda \left( 1 - \cos \theta \right) \right)} \\
\rho_{10} e^{t \left(i \omega - \lambda \left( 1 - \cos \theta \right) \right)} & \rho_{11}
\end{pmatrix}.
\end{equation}
From the latter result, we see that the system decoheres with rate $\lambda \left( 1 - \cos \theta \right)$.

\subsection{Finite number of ancillae}
Let us calculate the state of the system at time $t$ in the case in which there are $n$ ancillae. Given that the collision time of each of them is exponentially distributed with rate $\lambda/n$, the probability for any ancilla to have collided at time $t$ is $p_t = 1 - e^{-\lambda t/n}$. The convex sum over trajectories, equivalent to Eq.~\eqref{eq:rho_s_t_coll}, now reads
\begin{equation}\label{eq:convex_sum_over_num_colls_finite}
\begin{aligned}
\rho_\sys (t) &= \sum \limits_{\alpha_1=0}^{1} \cdots \sum \limits_{\alpha_n=0}^{1} \Lambda(\vec{\alpha}_{n}) \Phi_c^{\left(\sum_i \alpha_i\right)} \left[  U_t^\sys \rho_\sys (0) U_t^{\sys\dagger} \right].
\end{aligned}
\end{equation}
Each index $\alpha_i$ represents the state of the $i$-th ancilla, that is, whether it has collided or not, so the sum runs over all possible histories. The vector $\vec{\alpha}_{n} = (\alpha_1, \cdots, \alpha_n )$ has been introduced to compactify the notation. The function $\Lambda(\vec{\alpha}_{n})$ accounts for the probability of each trajectory, and is given by
\begin{equation}
\Lambda(\vec{\alpha}_{n}) = \prod\limits_{i=1}^{n} \left[ p_t^{\alpha_i} \left( 1 - p_t \right)^{1-\alpha_i} \right],
\end{equation}
Using again Eq.~\eqref{eq:k_collisions}, we obtain
\begin{equation}
\rho_\sys (t) = \begin{pmatrix}
\rho_{00} & \rho_{01} e^{-i \omega t} \left\langle (\cos\theta)^{\sum_i \alpha_i}  \right\rangle_{t}\\
\rho_{10} e^{i \omega t} \left\langle (\cos\theta)^{\sum_i \alpha_i} \right\rangle_{t} & \rho_{11}
\end{pmatrix},
\end{equation}
where $\langle \cdot \rangle_{t}$ stands for the average over trajectories at time $t$. This quantity can be computed as
\begin{equation}\label{eq:non-markov_dynam_map}
\begin{aligned}
\left\langle (\cos\theta)^{\sum_i \alpha_i} \right\rangle_{t} &= \sum \limits_{\alpha_1=0}^{1} \cdots \sum \limits_{\alpha_n=0}^{1} \Lambda(\vec{\alpha}_{n}) (\cos\theta)^{\sum_i \alpha_i}  \\
& = \sum \limits_{\alpha_1=0}^{1} \cdots \sum \limits_{\alpha_n=0}^{1} \prod\limits_{i=1}^{n} \left[ \left(p_t \cos \theta\right)^{\alpha_i} \left( 1 - p_t \right)^{1-\alpha_i} \right] \\
& = \left( p_t \cos \theta + 1 - p_t \right)^n = \left[ 1 + \left( \cos \theta -1 \right) \left(1 - e^{-\lambda t/n} \right) \right]^n.
\end{aligned}
\end{equation}
In this case, the entanglement-invariance in the dynamics no longer holds. Furthermore, this new phase factor $c_\mathrm{NM}(t)$ gives rise to non-Markovian dynamics (see main text)º.

\section{Quantum Darwinism}\label{sec:Darwinism}
In this section, we compute the reduced density operators of the system and a fraction of the environment, as well as their corresponding von Neumann entropies.

\subsection{Mutual information between system and ancillae}
The following calculations correspond to the situation in which the environment is composed by the ancillae only. As explained in the main text, we particularise to non-entangling interaction, $\theta = \pi$, as well as to $\rho_\sys (0) = | + \rangle \langle + |$ as the initial state of the system. The total system-environment state at time $t$, $\rho_{\sys\mathcal{E}}(t)$, can be written as the convex sum
\begin{equation}\label{eq:total_state_ancillae}
\begin{aligned}
\rho_{\sys\mathcal{E}}(t) &= \sum \limits_{\alpha_1=0}^{1} \cdots \sum \limits_{\alpha_n=0}^{1} \Lambda(\vec{\alpha}_{n}) | \vec{\alpha}_{n} \rangle \langle \vec{\alpha}_{n} |\otimes \left( \delta_{\mathrm{even}}(\vec{\alpha}_{n}) | +_t \rangle \langle +_t | + \delta_{\mathrm{odd}}(\vec{\alpha}_{n}) | -_t \rangle \langle -_t | \right),
\end{aligned}
\end{equation}
where, using the same notation from the previous section, $| \vec{\alpha}_{n} \rangle \equiv | \alpha_1 \cdots \alpha_n \rangle$. Furthermore, we have denoted $| \pm_t \rangle = U_t^\sys | \pm \rangle$ and defined the functions $ \delta_{\mathrm{even}}(\vec{\alpha}_{n})$ and $ \delta_{\mathrm{odd}}(\vec{\alpha}_{n})$, which are equal to one if their argument represents an even (odd) number of collisions and zero otherwise. These can be written as $ \delta_{\mathrm{even}}(\vec{\alpha}_{n}) = \frac{1 + (-1)^{\sum_i \alpha_i}}{2}$ and $ \delta_{\mathrm{odd}}(\vec{\alpha}_{n}) = \frac{1 - (-1)^{\sum_i \alpha_i}}{2}$. Equation \eqref{eq:total_state_ancillae} explicitly shows that the state is fully separable. Now, in order to compute the mutual information, we need to characterise the reduced state of the system and a fraction $f$ of the environment, $\rho_{\sys\mathcal{E}_f}(t) = \mathrm{Tr}_{k} \left[ \rho_{\sys\mathcal{E}}(t) \right]$, resulting from tracing out $k = (1-f)n$ ancillae. To do so, consider a projector $| \vec{\alpha}_{n-k} \rangle \langle \vec{\alpha}_{n-k} |$ corresponding to some state of the $n-k$ non-traced-out ancillae representing an even number of collisions. In the reduced density operator, it will appear tensored with both $| +_t \rangle \langle +_t |$ and $| -_t \rangle \langle -_t |$, since $| \vec{\alpha}_{n-k} \rangle \langle \vec{\alpha}_{n-k} | \otimes | +_t \rangle \langle +_t |$ will be the result of integrating over all the states of the $k$ traced-out ancillae representing an even number of collisions, whereas the integration over odd collisions of the $k$ ancillae will give $| \vec{\alpha}_{n-k} \rangle \langle \vec{\alpha}_{n-k} | \otimes | -_t \rangle \langle -_t |$. Now, their corresponding matrix elements can be readily computed, since they are simply given by the probability of $\vec{\alpha}_{n-k}$ multiplied by the sum of the probabilities of all the states of the $k$ traced-out ancillae with either even or odd parities. Hence, we can write $\rho_{\vec{\alpha}_{n-k},+_t} = \Lambda(\vec{\alpha}_{n-k}) P_{k}^{\mathrm{e}}(t)$ and $\rho_{\vec{\alpha}_{n-k},-_t} = \Lambda(\vec{\alpha}_{n-k}) \left( 1 - P_{k}^{\mathrm{e}}(t) \right)$, where $P_{k}^{\mathrm{e}}(t)$ is the probability for the $k$ ancillae to yield an even number of collisions at time $t$ and can be computed as
\begin{equation}
\begin{aligned}
P_{k}^{\mathrm{e}}(t) &= \sum \limits_{\alpha_1=0}^{1} \cdots \sum \limits_{\alpha_k=0}^{1} \Lambda(\vec{\alpha}_{k}) \delta_{\mathrm{even}}(\vec{\alpha}_{k}) \\
&= \frac{1}{2} \left(1 + \sum \limits_{\alpha_1=0}^{1} \cdots \sum \limits_{\alpha_k=0}^{1} \Lambda(\vec{\alpha}_{k}) (-1)^{\sum_i \alpha_i} \right)\\
&= \frac{1}{2} \left[1 + \left( 2e^{-\lambda t/n} -1 \right)^{k} \right].
\end{aligned}
\end{equation}
In the last step of the above calculation, we have proceeded as in Eq.~\eqref{eq:non-markov_dynam_map}. Notice that $+_t$ and $-_t$ need to be swapped in the previous discussion if the parity of $\vec{\alpha}_{n-k}$ is odd. With this result in hand, we can now compute all reduced density matrices in diagonal form. First, by setting $k = n$, we immediately obtain the reduced state of the system as $\rho_\sys (t) = P_{n}^{\mathrm{e}}(t) | +_t \rangle \langle +_t | + \left(1- P_{n}^{\mathrm{e}}(t) \right) | -_t \rangle \langle -_t |$, which agrees with our previous calculation, Eq.~\eqref{eq:non-markov_dynam_map}. The reduced state of the environment is given by $\rho_{\mathcal{E}_f}(t) = \mathrm{Tr}_\sys \left[ \rho_{\sys\mathcal{E}_f}(t) \right] = \sum \limits_{\alpha_1=0}^{1} \cdots \sum \limits_{\alpha_{n-k}=0}^{1} \left( \rho_{\vec{\alpha}_{n-k},+_t} + \rho_{\vec{\alpha}_{n-k},-_t} \right) | \vec{\alpha}_{n-k} \rangle \langle \vec{\alpha}_{n-k} | = \sum \limits_{\alpha_1=0}^{1} \cdots \sum \limits_{\alpha_{n-k}=0}^{1} \Lambda(\vec{\alpha}_{n-k})| \vec{\alpha}_{n-k} \rangle \langle \vec{\alpha}_{n-k} |$. With these results, we can immediately compute the entropies involved in the mutual information. We have
\begin{equation}
\begin{aligned}
H_\sys &= - P_{n}^{\mathrm{e}}(t) \log P_{n}^{\mathrm{e}}(t) - \left( 1 - P_{n}^{\mathrm{e}}(t) \right) \log \left( 1 - P_{n}^{\mathrm{e}}(t) \right),\\
H_{\mathcal{E}_f} &= -\sum \limits_{\alpha_1=0}^{1} \cdots \sum \limits_{\alpha_{n-k}=0}^{1} \Lambda(\vec{\alpha}_{n-k}) \log \left( \Lambda(\vec{\alpha}_{n-k}) \right),\\
H_{\sys\mathcal{E}_f} &= -\sum \limits_{\alpha_1=0}^{1} \cdots \sum \limits_{\alpha_{n-k}=0}^{1}  \sum \limits_{\alpha_\sys=0}^{1} \Lambda(\vec{\alpha}_{n-k}) \left[ P_{k}^{\mathrm{e}}(t)^{\alpha_\sys} \left( 1 - P_{k}^{\mathrm{e}}(t) \right)^{1-\alpha_\sys} \right] \\
& \times \log \left( \Lambda(\vec{\alpha}_{n-k}) \left[ P_{k}^{\mathrm{e}}(t)^{\alpha_\sys} \left( 1 - P_{k}^{\mathrm{e}}(t) \right)^{1-\alpha_\sys} \right] \right),
\end{aligned}
\end{equation}
where we have included an additional index $\alpha_\sys$ in the last expression to account for the state of the system. Given that both $H_{\mathcal{E}_f}$ and $H_{\sys\mathcal{E}_f}$ can be interpreted as the entropies of the joint probability distributions of independent random variables, they can be expressed as the sum of their individual entropies, so we can write $H_{\mathcal{E}_f}- H_{\sys\mathcal{E}_f} = P_{k}^{\mathrm{e}}(t) \log P_{k}^{\mathrm{e}}(t) + \left( 1 - P_{k}^{\mathrm{e}}(t) \right) \log \left( 1 - P_{k}^{\mathrm{e}}(t) \right)$, that is
\begin{equation}\label{eq:mutual_information_ancillae}
I_f = P_{k}^{\mathrm{e}}(t) \log P_{k}^{\mathrm{e}}(t) + \left( 1 - P_{k}^{\mathrm{e}}(t) \right) \log \left( 1 - P_{k}^{\mathrm{e}}(t) \right) - P_{n}^{\mathrm{e}}(t) \log P_{n}^{\mathrm{e}}(t) - \left( 1 - P_{n}^{\mathrm{e}}(t) \right) \log \left( 1 - P_{n}^{\mathrm{e}}(t) \right),
\end{equation}
where $f = 1-\frac{k}{n}$.

\subsection{The super-environment}
We now study the case in which the ancillae are emitted as a consequence of a quantum process. As explained in the main text, the emitter-ancilla dynamics is such that their joint state at time $t$ can be written as $\sqrt{1 - p_t} |1\rangle_e \otimes |0\rangle_a + \sqrt{p_t} |0\rangle_e \otimes |1\rangle_a$. Hence, the total state including system and environment is now the pure state
\begin{equation}
| \psi_{\sys\mathcal{E}}(t) \rangle = \sum \limits_{\alpha_1=0}^{1} \cdots \sum \limits_{\alpha_n=0}^{1} \sqrt{\Lambda(\vec{\alpha}_{n})} | \vec{\bar{\alpha}}_{n} \rangle_e \otimes | \vec{\alpha}_{n} \rangle_a \otimes \left( \delta_{\mathrm{even}}(\vec{\alpha}_{n}) | +_t \rangle + \delta_{\mathrm{odd}}(\vec{\alpha}_{n}) | -_t \rangle \right),
\end{equation}
where $\bar{\alpha}_i = 1 - \alpha_i$ represents the state of emitter $i$. Tracing out all the emitters cancels all the coherences in the density matrix and leaves us with Eq.~\eqref{eq:total_state_ancillae}. Therefore, despite the quantum super-environment, our previous result regarding the lack of information in the ancillae alone still holds. However, we are now interested in the mutual information as we trace out emitter-ancilla pairs. As we did in the previous subsection, we need the reduced state $\rho_{\sys\mathcal{E}_f}(t)$ resulting from the partial trace of $k$ pairs. Furthermore, since we must compute von Neumann entropies, our strategy consists in finding the orthogonal pure states whose convex sum leads to the reduced density operators; by doing so, the entropies are simply given by the Shannon entropies of the corresponding probabilities. To simplify the notation in what follows, let us define $| \beta_i \rangle \equiv | \bar{\alpha}_i \rangle_e \otimes | \alpha_i \rangle_a$, with $\beta_i = \alpha_i$, to account for the state of a pair. Now, to determine $\rho_{\sys\mathcal{E}_f}(t)$, we proceed in the following way. First, we notice that, when summing over the states of the $k$ ancillae $|  \vec{\beta}_k \rangle$, if $\vec{\beta}_k$ represents an even number of collisions, $\delta_{\mathrm{even}}(\vec{\beta}_{n-k}) = \delta_{\mathrm{even}}(\vec{\beta}_{n})$, so $\langle \vec{\beta}_k| \psi_{\sys\mathcal{E}}(t) \rangle \langle \psi_{\sys\mathcal{E}}(t) |  \vec{\beta}_k \rangle$ yields $\Lambda(\vec{\beta}_{k}) | \psi_{\sys\mathcal{E}_f,\mathrm{e}}(t) \rangle \langle \psi_{\sys\mathcal{E}_f,\mathrm{e}}(t) |$, where
\begin{equation}
| \psi_{\sys\mathcal{E}_f,\mathrm{e}}(t) \rangle = \sum \limits_{\beta_1=0}^{1} \cdots \sum \limits_{\beta_{n-k}=0}^{1} \sqrt{\Lambda(\vec{\beta}_{n-k})} | \vec{\beta}_{n-k} \rangle \otimes \left( \delta_{\mathrm{even}}(\vec{\beta}_{n-k}) | +_t \rangle + \delta_{\mathrm{odd}}(\vec{\beta}_{n-k}) | -_t \rangle \right).
\end{equation}
If $\vec{\beta}_k$ represents an odd number of collisions, the same operation yields $\Lambda(\vec{\beta}_{k}) | \psi_{\sys\mathcal{E}_f,\mathrm{o}}(t) \rangle \langle \psi_{\sys\mathcal{E}_f,\mathrm{o}}(t) |$ instead, with
\begin{equation}
| \psi_{\sys\mathcal{E}_f,\mathrm{o}}(t) \rangle = \sum \limits_{\beta_1=0}^{1} \cdots \sum \limits_{\beta_{n-k}=0}^{1} \sqrt{\Lambda(\vec{\beta}_{n-k})} | \vec{\beta}_{n-k} \rangle \otimes \left( \delta_{\mathrm{odd}}(\vec{\beta}_{n-k}) | +_t \rangle + \delta_{\mathrm{even}}(\vec{\beta}_{n-k}) | -_t \rangle \right).
\end{equation}
Since we must sum over all $| \vec{\beta}_{k} \rangle$, the resulting state can be written as
\begin{equation}\label{eq:state_system_fraction_pairs}
\rho_{\sys\mathcal{E}_f}(t) = P_{k}^{\mathrm{e}}(t) | \psi_{\sys\mathcal{E}_f,\mathrm{e}}(t) \rangle \langle \psi_{\sys\mathcal{E}_f,\mathrm{e}}(t) | + \left( 1 - P_{k}^{\mathrm{e}}(t) \right) | \psi_{\sys\mathcal{E}_f,\mathrm{o}}(t) \rangle \langle \psi_{\sys\mathcal{E}_f,\mathrm{o}}(t) |.
\end{equation}
From this expression, we can immediately see that the reduced state of the system is, as in the case without emitters, $\rho_\sys (t) = P_{n}^{\mathrm{e}}(t) | +_t \rangle \langle +_t | + \left(1- P_{n}^{\mathrm{e}}(t) \right) | -_t \rangle \langle -_t |$. As for the reduced state of the environment, $\rho_{\mathcal{E}_f}(t)$, we observe that $\langle +_t | \psi_{\sys\mathcal{E}_f,\mathrm{e}}(t) \rangle \langle \psi_{\sys\mathcal{E}_f,\mathrm{e}}(t) | +_t \rangle = \langle -_t | \psi_{\sys\mathcal{E}_f,\mathrm{o}}(t) \rangle \langle \psi_{\sys\mathcal{E}_f,\mathrm{o}}(t) | -_t \rangle = P_{n-k}^{\mathrm{e}}(t) | \psi_{\mathcal{E}_f,\mathrm{e}}(t) \rangle \langle \psi_{\mathcal{E}_f,\mathrm{e}}(t) |$, where
\begin{equation}
| \psi_{\mathcal{E}_f,\mathrm{e}}(t) \rangle = \sum \limits_{\left\lbrace \vec{\beta}_{n-k} | \delta_\mathrm{even}(\vec{\beta}_{n-k}) = 1 \right\rbrace} \sqrt{ \frac{\Lambda(\vec{\beta}_{n-k})}{P_{n-k}^{\mathrm{e}}(t)}} | \vec{\beta}_{n-k} \rangle.
\end{equation}
Similarly, $\langle -_t | \psi_{\sys\mathcal{E}_f,\mathrm{e}}(t) \rangle \langle \psi_{\sys\mathcal{E}_f,\mathrm{e}}(t) | -_t \rangle = \langle +_t | \psi_{\sys\mathcal{E}_f,\mathrm{o}}(t) \rangle \langle \psi_{\sys\mathcal{E}_f,\mathrm{o}}(t) | +_t \rangle = \left( 1 - P_{n-k}^{\mathrm{e}}(t) \right) | \psi_{\mathcal{E}_f,\mathrm{o}}(t) \rangle \langle \psi_{\mathcal{E}_f,\mathrm{o}}(t) |$, where
\begin{equation}
| \psi_{\mathcal{E}_f,\mathrm{o}}(t) \rangle = \sum \limits_{\left\lbrace \vec{\beta}_{n-k} | \delta_\mathrm{odd}(\vec{\beta}_{n-k}) = 1 \right\rbrace} \sqrt{ \frac{\Lambda(\vec{\beta}_{n-k})}{1 - P_{n-k}^{\mathrm{e}}(t)}} | \vec{\beta}_{n-k} \rangle.
\end{equation}
Finally, using Eq.~\eqref{eq:state_system_fraction_pairs}, we obtain
\begin{equation}
\begin{aligned}
\rho_{\mathcal{E}_f}(t) & = P_{k}^{\mathrm{e}}(t) \left( P_{n-k}^{\mathrm{e}}(t) | \psi_{\mathcal{E}_f,\mathrm{e}}(t) \rangle \langle \psi_{\mathcal{E}_f,\mathrm{e}}(t) | + \left( 1 - P_{n-k}^{\mathrm{e}}(t) \right) | \psi_{\mathcal{E}_f,\mathrm{o}}(t) \rangle \langle \psi_{\mathcal{E}_f,\mathrm{o}}(t) | \right) \\
&+ \left( 1 - P_{k}^{\mathrm{e}}(t) \right) \left( \left( 1 - P_{n-k}^{\mathrm{e}}(t) \right) | \psi_{\mathcal{E}_f,\mathrm{o}}(t) \rangle \langle \psi_{\mathcal{E}_f,\mathrm{o}}(t) | + P_{n-k}^{\mathrm{e}}(t) | \psi_{\mathcal{E}_f,\mathrm{e}}(t) \rangle \langle \psi_{\mathcal{E}_f,\mathrm{e}}(t) | \right)\\
&= P_{n-k}^{\mathrm{e}}(t) | \psi_{\mathcal{E}_f,\mathrm{e}}(t) \rangle \langle \psi_{\mathcal{E}_f,\mathrm{e}}(t) | + \left( 1 - P_{n-k}^{\mathrm{e}}(t) \right) | \psi_{\mathcal{E}_f,\mathrm{o}}(t) \rangle \langle \psi_{\mathcal{E}_f,\mathrm{o}}(t) |.
\end{aligned}
\end{equation}
Hence, in this case, we have
\begin{equation} 
\begin{aligned}
H_\sys &= - P_{n}^{\mathrm{e}}(t) \log P_{n}^{\mathrm{e}}(t) - \left( 1 - P_{n}^{\mathrm{e}}(t) \right) \log \left( 1 - P_{n}^{\mathrm{e}}(t) \right),\\
H_{\mathcal{E}_f} &= - P_{n-k}^{\mathrm{e}}(t) \log P_{n-k}^{\mathrm{e}}(t) - \left( 1 - P_{n-k}^{\mathrm{e}}(t) \right) \log \left( 1 - P_{n-k}^{\mathrm{e}}(t) \right),\\
H_{\sys\mathcal{E}_f} &= - P_{k}^{\mathrm{e}}(t) \log P_{k}^{\mathrm{e}}(t) - \left( 1 - P_{k}^{\mathrm{e}}(t) \right) \log \left( 1 - P_{k}^{\mathrm{e}}(t) \right),
\end{aligned}
\end{equation}
From this, the mutual information follows,
\begin{equation}\label{eq:mutual_information_emitters}
\begin{aligned}
I_f &= P_{k}^{\mathrm{e}}(t) \log P_{k}^{\mathrm{e}}(t) + \left( 1 - P_{k}^{\mathrm{e}}(t) \right) \log \left( 1 - P_{k}^{\mathrm{e}}(t) \right) \\
&- P_{n}^{\mathrm{e}}(t) \log P_{n}^{\mathrm{e}}(t) - \left( 1 - P_{n}^{\mathrm{e}}(t) \right) \log \left( 1 - P_{n}^{\mathrm{e}}(t) \right) \\
&- P_{n-k}^{\mathrm{e}}(t) \log P_{n-k}^{\mathrm{e}}(t) - \left( 1 - P_{n-k}^{\mathrm{e}}(t) \right) \log \left( 1 - P_{n-k}^{\mathrm{e}}(t) \right).
\end{aligned}
\end{equation}
Comparing this result with Eq.~\eqref{eq:mutual_information_ancillae}, we see that the difference between both is the entropy of the environment, $H_{\mathcal{E}_f}$.

\end{document}